# USE OF GPS NETWORK DATA
# FOR HF DOPPLER MEASUREMENTS INTERPRETATION

**I. R. PETROVA, V. V. BOCHKAREV, R. R. LATYPOV**

Kazan Federal University, 18 Kremlevskaya Str., Kazan 42008, Russia
Inna.petrova@kpfu.ru

**Abstract.** The method of measurement of Doppler frequency shift of ionospheric signal – HF Doppler technique – is one of well-known and widely used methods of ionosphere research. It allows to research various disturbances in the ionosphere. There are some sources of disturbances in the ionosphere. These are geomagnetic storms, solar flashes, metrological effects, atmospheric waves. This method allows to find out the influence of earthquakes, explosions and other processes on the ionosphere, which occur near to the Earth. HF Doppler technique has the high sensitivity to small frequency variations and the high time resolution, but interpretation of results is difficult. In this work we make an attempt to use GPS data for Doppler measurements interpretation. Modeling of Doppler frequency shift variations with use of TEC allows to separate ionosphere disturbances of medium scale.

## 1. Introduction

Study of processes on upper atmosphere and ionosphere is one of the most important problems of geophysics. HF Doppler technique based on frequency shift measurements is widely used for ionosphere investigation. Advantages of the HF Doppler technique are the high sensitivity to small frequency variations and the high time resolution. It allows to research various disturbances in the ionosphere.

This technique is used at radiophysics department of Kazan University (Russia) since 2001 for ionosphere sounding. The measuring equipment is a system of closely spaced receivers. Characteristics of the measuring equipment, technique of experiment and data processing are described in [Petrova et al., 2007]. HF signals of two radio stations have been received. They are *Radio Rossii* (Radio of Russia) and special radio station of exact time. The transmitter of *Radio Rossii* radio station is located at Arkhangelsk (67°55'N, 33°01'E), the frequency is 6160 kHz, the transmitter of special radio station is located at Moscow (55°45'N, 37°18'E) and the frequencies are 4996 and 9996 kHz. The receivers are located at Kazan (55°49'N, 49°08'E).
Doppler measurements allow to analyze the time properties of ionospheric disturbances – the waves periods extend in the range from seconds to days. The spectral analysis gives information about typical periods of wave processes and allows to detect ionospheric irregularities of different scales. The results of spectral analysis for Doppler shift variations with periods in range from one minute to 60 days have been presented in [Petrova et al., 2009].

But the Doppler shift is an integral characteristic and includes information on all the variations of electron content along the propagation path of radio waves. Only Doppler measurements don't provide information about localization and origin of the electron density variations. To analyze the influence of different factors, we tried to calculate the value of Doppler shift using the total electron content (TEC) data.

The use of the international ground-based network of two-frequency receivers of the navigation GPS system makes possible a global, continuous and fully computerized monitoring of ionospheric disturbances of a different class. One of the applications of GPS-data for ionosphere research is Global Ionospheric Maps (GIM). The GIM-technique allows to produce global maps of absolute vertical value of TEC using GPS data [Mannucci et al.,1998, Schaer et al.,1998].

We used the GIM-data to model the Doppler shift. The GIM-data were used for the following reasons [Petrova et al., 2010]. Firstly, these data are easy of access. Secondly, they allow to estimate only the large-scale regular variations. This simplifies the procedure for Doppler shift calculating and, at a later, this estimation can be used to separate the influence of regular variations and disturbances. In this work we make an attempt to use GPS data for modeling of Doppler shift variations.

**2. Model description**

Variations of the ionosphere electron density influence both the Doppler frequency shift (DFS), and the TEC. We suppose that there is a correlation between these values. Regression analysis allows to model and analyze relationship between a dependent variable and one or more independent variables [Freedman, 2005]. In our work we use the linear regression model. As is known Doppler shift is dependent on the TEC changes along the radio wave path propagation. Four parameters were selected for modeling of Doppler shift:

$\frac{\partial I}{\partial t}$ - time derivative of TEC;

$\frac{\partial I}{\partial \varphi}$ - latitude derivative of TEC;

$\frac{\partial I}{\partial \theta}$ - longitude derivative of TEC;

$I$ - value of TEC.

Time derivative of TEC is related to variations of the vertical electron content. Spatial derivatives determine the trajectory of radio waves propagation. In this case the equation of linear regression has the form

$$\Delta \hat{f} = a_1 I + a_2 \frac{\partial I}{\partial t} + a_3 \frac{\partial I}{\partial \varphi} + a_4 \frac{\partial I}{\partial \theta} + a_5 \qquad (1)$$

The last parameter in the model determines the systematic error. The next step of regression analysis is to determine the values of model coefficients $a_1, a_2, a_3, a_4, a_5$ using experimental data. The system of equations for the model coefficients can be represented in matrix form:

$$\begin{bmatrix} \Delta f_1 \\ \Delta f_2 \\ \ldots \\ \Delta f_n \end{bmatrix} = \begin{bmatrix} I_1 & \left.\frac{\partial I}{\partial t}\right|_1 & \left.\frac{\partial I}{\partial \varphi}\right|_1 & \left.\frac{\partial I}{\partial \theta}\right|_1 & 1 \\ I_2 & \left.\frac{\partial I}{\partial t}\right|_2 & \left.\frac{\partial I}{\partial \varphi}\right|_2 & \left.\frac{\partial I}{\partial \theta}\right|_2 & 1 \\ \ldots & \ldots & \ldots & \ldots & \ldots \\ I_n & \left.\frac{\partial I}{\partial t}\right|_n & \left.\frac{\partial I}{\partial \varphi}\right|_n & \left.\frac{\partial I}{\partial \theta}\right|_n & 1 \end{bmatrix} \times \begin{bmatrix} a_1 \\ a_2 \\ a_3 \\ a_4 \\ a_5 \end{bmatrix} \qquad (2)$$

And the more compact form is $y = Xa$, **X** - matrix of system, **y** - vector of variables and **a** - vector of values for system coefficients.

We can use the least-squares method for solution of such system. This method is a standard approach to the approximate solution of overdetermined systems, i.e. sets of equations in which there are more equations than unknowns [Anton, 2005]. The most important application is in data fitting. The best fit in the least-squares sense minimizes the sum of squared residuals, a residual being the difference between an observed value and the fitted value provided by a model.

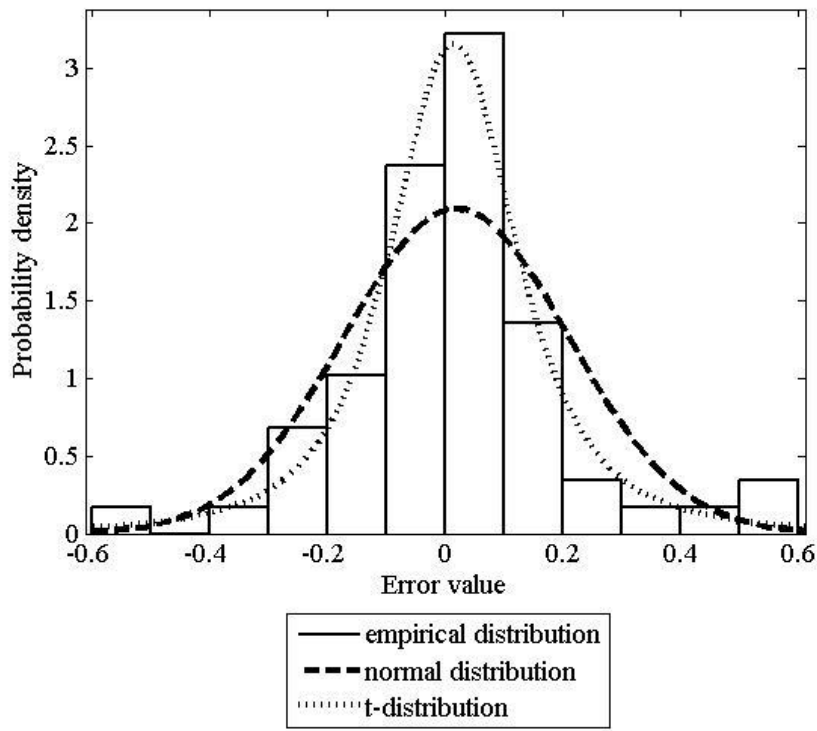

Figure 1. Experimental errors distribution and theoretical approximation

This algorithm of least-squares estimation (LSE) for calculating the model coefficients is simple but has limitations. It gives unsatisfactory results when the error distribution is different from the normal law. Figure 1 shows the experimental errors distribution and two theoretical curves: normal law and Student t-distribution. Normal law is not suited for experimental distribution. Approximation with help Student t-distribution is a more correct. Therefore the criterion of least squares is not desirable to calculate the coefficients of this model.

## 3. Estimation of model coefficient by the maximum likelihood criterion

If the errors distribution is known, then we can calculate the model coefficients by the maximum likelihood method. Maximum likelihood estimation (MLE) is a popular statistical method used for fitting a statistical model to data, and providing estimates for the model's parameters [Hahn et al., 1994]. In this case we suppose that the logarithmic likelihood function is maximal and calculate model coefficient for this conditions. The following iterative procedure is used for computing.

- Calculate the model coefficients by the least squares criterion.
- Calculate the DFS values according to the model.
- Determine the errors distribution.
- Calculate the model coefficients by the maximum likelihood criterion.
- If change of model coefficients is small then the calculation is completed.
- Else the calculation is continued with new coefficients.

The simulation results for the linear regression model are presented in Figure 2.

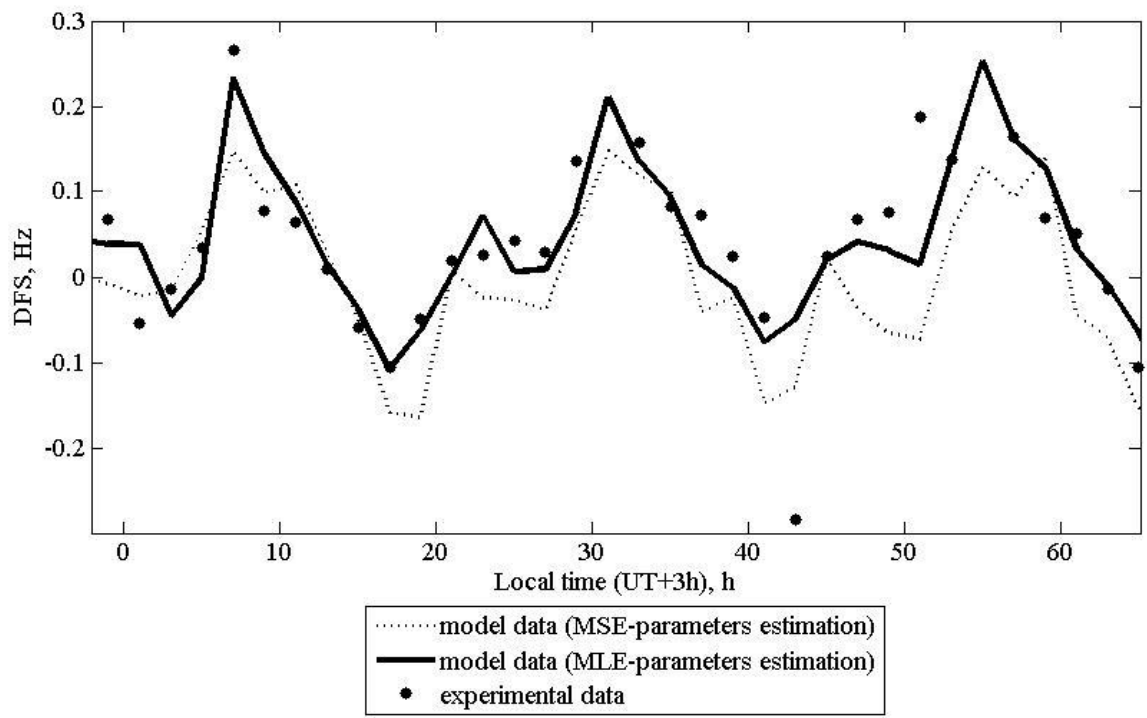

Figure 2. Simulation result for linear regression model.
Using experimental data: 4-6 March, 2006; frequency - 4996 kHz; Moscow – Kazan.

Relative error of model for LSE is 0.7 and for MLE is 0.5. Coefficient of determination for LSE is 0.5 and for MLE is 0.7 – 0.75.

## 4. Adaptive model of linear regression

Ionospheric parameters and conditions vary with time. An adequate model must take into account these changes and update the coefficients. In such cases adaptive models can be used. Usually adaptive models are applied for short-term forecast [Widrow, 1985].

Initially we have current values of model parameters and predict value for next time interval. The difference between the predicted and experimental values is calculated and used for correction of model coefficients. The model with new coefficients and experimental data is in better agreement. Than next value is predicted and update procedure for model coefficient is repeat. The aim of this "model training" is selection of optimal values for model coefficients.

The algorithm of adaptive procedure for calculating the model coefficients is presented below.

- Estimate the covariance matrix for initial data segment:
  $B = X_{init}^H X_{init}$, $R = X_{init}^H Y_{init}$
- Calculate the model coefficients for initial data segment: $a = (B)^{-1} R$
- Take next data segment: $X_{next}, Y_{next}$
- Update the covariance matrix and the model coefficients:
  $B^{new} = (1-\alpha)B + \alpha X_{next}^H X_{next}$, $R^{new} = (1-\alpha)R + \alpha X_{next}^H Y_{next}$, $a^{new} = (B^{new})^{-1} R^{new}$

B - is the covariance matrix of the parameters included in the model. R- covariance vector of Doppler shift and parameters. Initial segment of data is selected and the covariance matrix and vector are calculated. Then the initial values of the model coefficient are calculated. Alpha parameter determines the model update rate. If this parameter is small then the model slowly adapts to changing conditions. If the parameter is large then the model is very sensitive to random variations in the input data. This parameter has an optimum value. Prediction error was analyzed to select the optimal value alpha. Figure 3 shows the dependence of prediction error on alpha-coefficient. The value of alpha-coefficient for minimum error is 0.063. The life time of model is about 30 hours. Figure 4 shows the simulation results for adaptive model. The relative prediction error is 0.43.

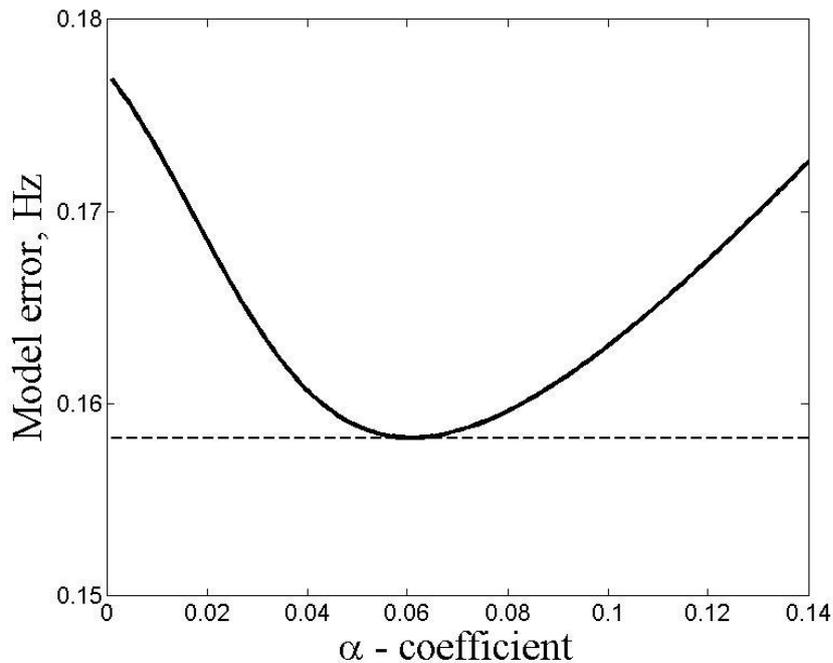

Figure 3. Dependence of prediction error on alpha-coefficient.

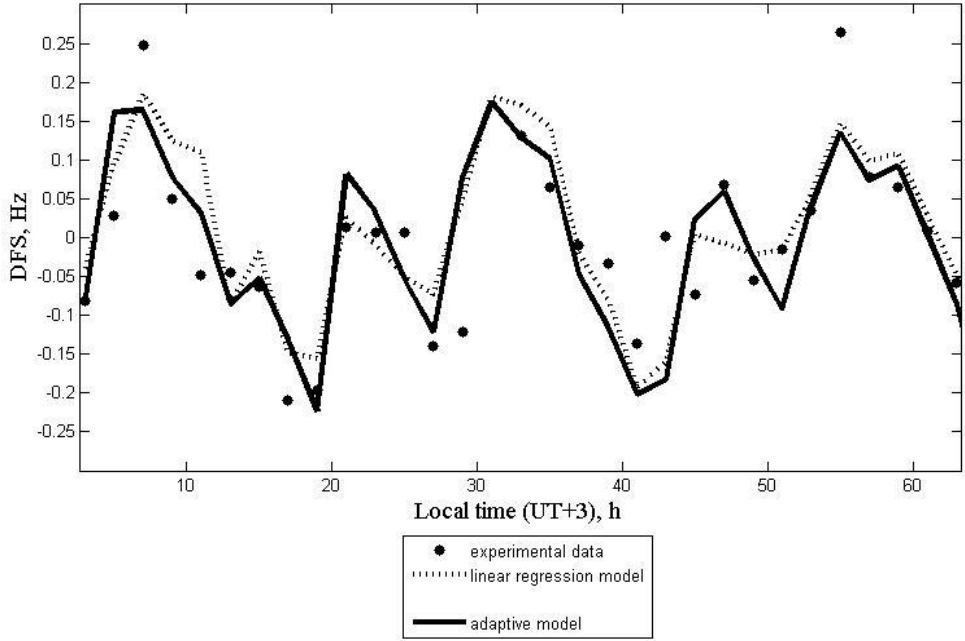

Figure 4. Simulation result for adaptive model.
Using experimental data: 2-3 January, 2010; frequency - 4996 kHz; Moscow – Kazan.

## 5. Relative contribution of model parameters to Doppler shift estimation

The model uses four parameters affecting the Doppler shift in different ways: TEC value, time derivative and two spatial derivatives – latitudinal and longitudinal. It is of interest to estimate the impact of each parameter of the model. The equation (1) can be presented as $\Delta \hat{f} = a_1 F_1 + a_2 F_2 + a_3 F_3 + a_4 F_4 + a_5$, where $F_i$ are parameters $I, \frac{\partial I}{\partial t}, \frac{\partial I}{\partial \varphi}, \frac{\partial I}{\partial \theta}$. Then the dispersion of i-th term is $\sigma^2 = a_i^2 DF_i$, where $DF_i$ is dispersion of $F_i$.

The estimations of parameter contributions in the model value were made for the two radio paths: Moscow – Kazan and Arkhangelsk – Kazan (Figure 5A,5B). Time derivative gives the largest contribution for both cases. Observed radio paths have different orientation. One of them (Moscow – Kazan) is the zonal and the other (Arkhangelsk – Kazan) is the meridional oriented. The contribution of the transverse gradient is greater than the longitudinal in spite of path orientation.

## 6. The trend removal in the daily Doppler shift variations

The variations related with regular changes of electron concentration can have the high amplitude. As it shown by Petrova et al. [2007] Doppler shift of HF can reach 1-2 Hz in quiet geomagnetic conditions. It complicates selection of quasi-periodic variations related with ionosphere wave process and can cause errors of periods and amplitudes calculation for ionospheric disturbances. It is the most essential for investigation of disturbances with periods from minutes to hour. Model presented in this paper allows to exclude a trend caused by daily change of electronic concentration.

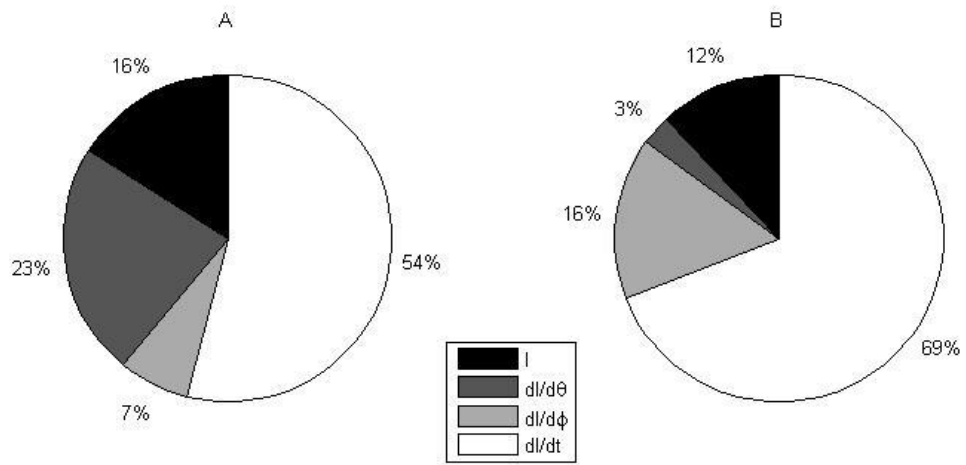

Figure 5. Relative contribution of model parameters to Doppler shift estimation.
A: N-S oriented radio path Archangelsk – Kazan, frequency 6160 kHz.
B: W-E oriented radio path Moscow– Kazan, frequency 4996 kHz.

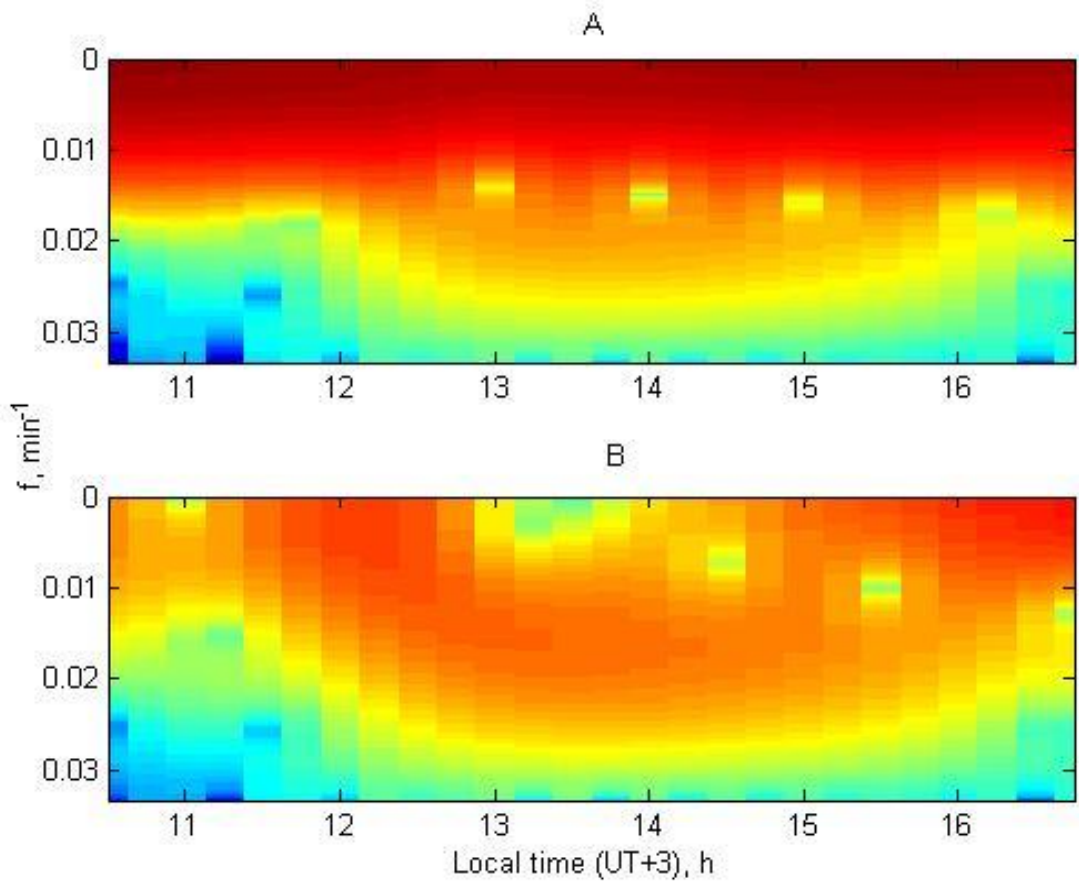

Figure 6. Doppler shift spectrogram. Using experimental data: April 24, 2006; frequency - 4996 kHz; Moscow – Kazan.
A: For Doppler shift series with daily trend. B: For Doppler shift series without daily trend.

Figure 6A shows the Doppler shift spectrogram. Fourier transform with Kaiser window has been used. Width of time window is 3 hours. We can see low-frequency component caused by the Doppler shift trend. Figure 6B shows the spectrogram for Doppler shift series without daily trend. This trend is caused by daily TEC variations and has been calculated on the basis of statistical model of DFS. Variations with the periods about an hour are visible in the range of 12-16 hours of local time. In Figure 6A such variations are not visible because of the low-frequency component.

Duration (number of wave cycles) of wave disturbances with the period from tens minutes to hours as a rule is not so much. In such cases the spectral methods of the high resolution (for example AR-method, MUSIC) are used for more exact definition of quasi-periodic variations parameters. As it is given by Marple [1987] these methods are sensitive to the trend in the analyzed time series. The trend removal is necessary procedure for application of such methods. The spectrum calculated by the MUSIC method is presented in Figure 7. The solid line is a spectrum calculated on an initial DFS series. Components with the periods less than 1.5 hours are not found out. Spectral maxima for periods in range of 2-3 hours are unstable and depend on the model order. The dashed line is a spectrum calculated on a series after the trend removal. In this case components with the periods about 1 hour are well separated.

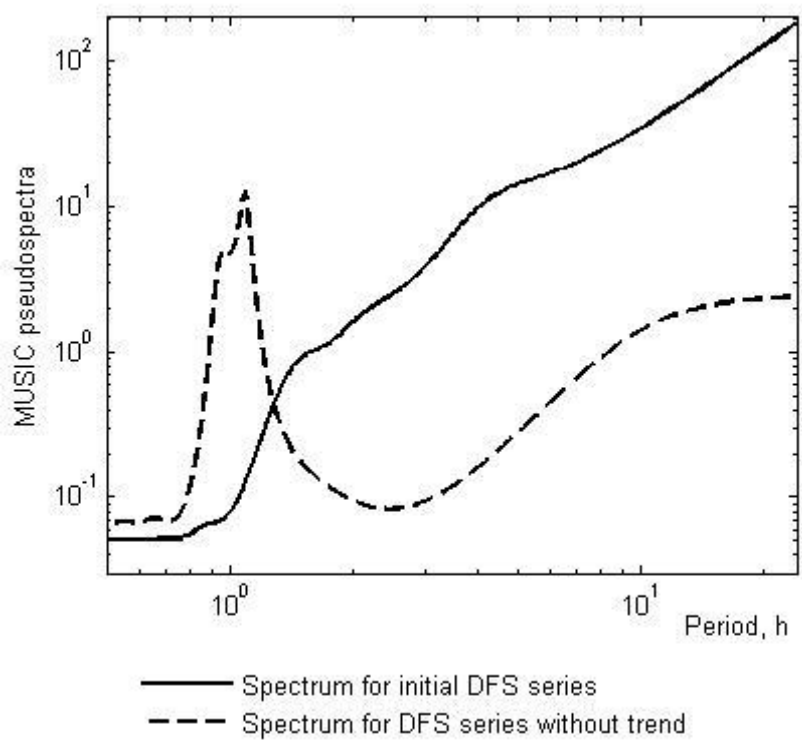

Figure 7. Doppler shift spectrum. Using experimental data: April 24, 2006; frequency - 4996 kHz; Moscow – Kazan.

### 6. Conclusions

In this paper we estimated the value of the Doppler shift using global ionospheric maps data. The linear regression model was applied. Model parameters are the values of total electron content and its derivatives. Different techniques for calculating the model coefficients were examined. Adaptive algorithm gives the best result.

Relative contribution of model parameters to Doppler shift estimation has been analyzed. Time derivative gives the largest contribution. We analyzed the data for the two differently oriented radio paths. It was found that the transverse gradient of the total electron content gives a greater contribution than the longitudinal for both radio paths.

Statistical model of Doppler shift has been used for trend removal. This procedure allows to separate spectral components efficiently.

**Acknowledgements**

We acknowledge the Crustal Dynamics Data Information System (CDDIS) for providing GPS data used in this study.